\documentclass[twocolumn,prl,preprintnumbers,amsmath,amssymb]{revtex4}

\usepackage{setspace}
\usepackage{graphicx}
\usepackage{dcolumn}
\usepackage{bm}
\usepackage{natbib}

\begin{document}
\title{Comment on "Evidence for a non Fermi liquid phase in Ge-substituted YbRh$_2$Si$_2$"}

\author{J. Gaddy, T. Heitmann, J. Lamsal,
and W. Montfrooij}
\affiliation{Department of Physics and the Missouri Research
Reactor, University of Missouri, Columbia, MO 65211}
\maketitle
In a recent paper, Custers {\it et al.} \cite{custers} argue for the existence of a new metallic quantum critical phase at 0 K in the Ge-doped heavy-fermion system YbRh$_2$Si$_2$ in the presence of magnetic frustration. In here we discuss the consequences of this identification for the (more standard) field induced quantum critical phase.\\
YbRh$_2$Si$_2$ is characterized by two $B$-dependent energy scales: the antiferromagnetic (AF) ordering temperature $T_N(B)$, and $T^*(B)$ below which non Fermi liquid behavior starts to evolve into Fermi liquid behavior.
Through doping \cite{custers,si,rh} and pressure \cite{pres} studies it has been established that $T_N$ is strongly dependent on (chemical) pressure, while $T^*$ is virtually independent of it.
In most quantum critical systems only one energy scale exists as both the strength of the RKKY interaction and the Kondo shielding temperature depend on orbital overlap. However, in a system where frustration has weakened the ordering tendency two energy scales should be present: $T_N(B)$ below which thermal fluctuations are overcome by the intermoment interaction, and $T^*(B)$ below which local moments anti-parallel to the external field are preferentially Kondo-shielded. Custers {\it et al.} \cite{custers} have already made the case for frustration in this stoichiometric compound, in here we argue that this decoupling of energy scales also implies that the phase for $T<T^*$ and $B>B_c$ is dominated by isolated magnetic clusters that originate in Kondo shielding and that behave very similar to heavily doped quantum critical systems.\\
Experiments identify $T^*$ as the Kondo shielding energy scale. First, Hall effect measurements \cite{hall} report an increase in Fermi surface on increasing $B$ through $B_c$, indicative of Kondo shielding taking place. It has already been argued \cite{moments} that the critical physics near $B=B_c$ originates from local criticality. Second, the demise of $T_N$ with increased $B$ is a consequence of raising the energy of the AF-ground state in a field; it cannot be viewed as a result of Kondo shielding as this would have resulted in an increase of the Fermi surface at $T_N(B)$=0 rather than at $T^*(B_c)$ \cite{hall}. Also, it would have implied that $T_N(B)$=0 would occur at $B=B_c$. Thus, $T^*$ is the average temperature below which anti-parallel moments are preferentially shielded. Note that such a scale should exist- in principle- in a system subject to Kondo-shielding.\\
In locally critical systems close to $T^*$  magnetic clusters can form through quantum fluctuations that randomly remove anti-parallel moments. These clusters consist of linked moments surrounded by shielded moments. For example, a cluster of 4 members with one central moment AF coupled to three others yields a cluster with a net superspin of 3-1=2. The clusters themselves are not intrinsically stable, unlike the case for heavily doped systems such as quantum critical CeRu$_{0.5}$Fe$_{1.5}$Ge$_2$ where such clusters have been observed \cite{w} in single crystal neutron scattering experiments. However, a sufficiently strong $B$-field stabilizes AF-ordered clusters with a net superspin. Of course, whether or not the moments within such a cluster do line up depends on the size of the cluster and on $T$. Finite size effects will force the moments in small clusters to line up at relatively high $T$, whereas larger clusters can remain unaligned down to low $T$. Increasing $B$ will favor increasingly smaller clusters, until at very high $B$ only isolated moments remain that follow the field rather than the AF-coupling tendency with their neighbors. We capture this in a field-dependent equilibrium cluster size $S(B)$.\\
Not only would such field-stabilized clusters neatly account for the FM response in the region $T_{FL}<T<T^*$ and $B>B_c$, their presence also explains the measured susceptibility $\chi (B,T)$. Cooling down from $T>T^*$ for $B>B_c$ we expect the following. First, small clusters with net moments will form, and these clusters will turn into superspins at a temperature dictated by their size. Once aligned, these clusters persist upon cooling as Kondo shielding is impeded by the local ordering. $\chi$ will increase during the approach of $T^*$ as larger clusters turn into superspins and the contributions of these emerging clusters will be added to those of smaller clusters. By the same token, $\chi$ should have the same $B$-independent value for all $T>T^*$ and $B>B_c$ as the $T$-dependence is governed entirely by the size distribution of the clusters. Once the equilibrium size $S(B)$ is reached (at $T=T^*$), $\chi$ will not increase any more; rather, for $T<T^*$ we will see the effects of dissipation. Dissipation  forces the larger clusters to choose a superspin direction, while the smaller clusters that can still follow the external field will determine $\chi(B,T=0)$. The closer we get to $B_c$, the larger $S(B)$ and hence the lower $T^*(B)$ at which the largest clusters change into stable superspins; the maximum value of $\chi$ at $T^*$ increases with decreasing $B$ because of the larger average cluster size. All this is exactly as observed (Fig. 1 in \cite{si}), making it very likely that stoichiometric quantum critical systems with magnetic frustration harbor clusters very similar to those observed \cite{w} in heavily doped systems.\\
This research is supported by the U.S. Department of Energy, Basic Energy Sciences, and the Division of Materials Sciences and Engineering under Grant No. DE-FG02-07ER46381.\\

\end{document}